\documentclass{aa}  

\renewcommand{\makeLineNumber}{}
\usepackage{graphicx}

\usepackage{txfonts}
\usepackage{bm}
\usepackage{xcolor}
\usepackage{amsmath}
\usepackage{graphicx}
\usepackage{mathtools} 
\usepackage{siunitx}
\newcommand{\nablab}{\ensuremath{\vec{\nabla}}}
\newcommand{\Ek}{\ensuremath{{\rm{Ek}}}}
\newcommand{\Rat}{\ensuremath{{\rm{Ra}}_T}}
\newcommand{\Rac}{\ensuremath{{\rm{Ra}}_C}}
\newcommand{\Pm}{\ensuremath{{\rm{Pm}}}}
\newcommand{\Sc}{\ensuremath{{\rm{Sc}}}}

\newcommand{\Rey}{\ensuremath{{\rm{Re}}}}
\newcommand{\Rm}{\ensuremath{{\rm{R}}_m}}
\begin{document}

\title{Planetary dynamos driven by semiconvection in stratified layers}

\author{P. Pru\v{z}ina \and D. C\'ebron \and N. Schaeffer}
\institute{ISTerre, Universit\'e Grenoble Alpes, 38610 Gi\`eres, France}

\date{Received date /
Accepted date }

\abstract{Stably stratified fluid layers are common in gaseous planets, stellar interiors, and planetary cores, and have long been considered incapable of sustaining dynamo action. Here, we show that semiconvection—driven by a destabilizing thermal gradient within an overall stably stratified medium—can, in fact, give rise to self-sustained magnetic fields. Motivated by recent models suggesting that large portions of Jupiter and Saturn may be semiconvective, we perform direct numerical simulations in spherical shells, operating in the planetary-relevant regime of low magnetic Prandtl numbers. From a primary semiconvection instability, a layered convection state spontaneously develops, consisting of a convective region beneath a stably stratified layer of comparable thickness. Fluid motions in this convective region are strong enough to produce magnetic fields with key features observed in planetary dynamos, including strong dipolarity, realistic field strengths, and spectral characteristics.  These results provide the first direct evidence that semiconvection can drive dynamo action in stably stratified regions of gas giants and stellar interiors, with important implications for understanding astrophysical magnetic field generation.}


\maketitle

\section{Introduction}\label{sec:introduction}

In the past few years, the Juno and Cassini missions have provided us with measurements of the magnetic fields of Jupiter and Saturn in high details, revealing complex fields to unprecedented resolution \citep{connerney2022new,dougherty2018saturn}. However, the processes supporting these fields are still not fully understood, with these studies showing surprising features that are difficult to explain with our current dynamo paradigms.
The gas giants are thought to be comprised of a surface layer of molecular hydrogen, above a deep core of metallic hydrogen, so-called for its high electrical conductivity. Planetary magnetic fields are likely maintained by the dynamo action of fluid flows in this metallic hydrogen core. The transition between these regions is gradual, but peak conductivity is reached by radius $80\%$ of the planetary radius for Jupiter, and $50\%$ in Saturn. \citep{jones2011planetary}.

Previous dynamo models have focused on overturning convection \citep[e.g.][]{gastine2012dipolar,jones2014dynamo}, but  convective layers in planets may be too deep inside the planetary interior to explain the features of the observed magnetic fields \citep[e.g.][]{debras2019new}. Recent structure models propose that, closer to the surface,  there may exist large regions with a stably stratified density due to the concentration of helium in the metallic hydrogen, but destabilising temperature gradients \citep{leconte2012new,debras2019new}. These regions are unstable to semiconvection, a form of double diffusive convection. This is a phenomenon in which fluid motions are driven by differences in the gradients of two scalars that both contribute to the fluid density, that diffuse at different rates \citep{schwarzschild1958evolution}. In astrophysical applications, the fast-diffusing scalar is temperature $T$, and the slow-diffusing scalar is composition $C$ of some heavy element (e.g. helium). However, semiconvection also occurs in terrestrial oceans, where salinity is the slow-diffusing component, and has been observed in laboratory experiments of salt-sugar solutions. The semiconvection regime is characterised by a destabilising thermal gradient, and stabilising compositional gradient; the opposite case is referred to as fingering convection.
An excellent introduction to the subject is provided by \citet{radko2013double}.

While it has not been strictly demonstrated---and we will observe that it does not hold in our study---, it is thought that the radial extent of dynamo action in a planet can be estimated based on the magnetic energy spectrum \citep{lowes1974spatial}. Using this approach for Jupiter, with the data from the Juno mission, gives an estimate of $r_{\text{dyn}}\approx 0.8R_J$, where $R_J$ is the radius of Jupiter \citep{connerney2022new}. However, based on the flow strength required to generate a dynamo, \citet{duarte2018physical} suggest a much shallower estimate, with the dynamo layer extending to $0.9R_J$.  For either estimate, it is likely that at least part of the dynamo region is semiconvective, with a stabilising composition gradient and destabilising temperature gradient \citep{leconte2012new,leconte2013layered,debras2019new}. However, recent work on dynamo simulations of Jupiter's magnetic field highlight that \textit{'dynamics of semi-convection in astrophysical regimes are poorly understood'}, and notably that \textit{'an investigation of the possibility of magnetic field generation under semi-convection is required'} \citep{moore2022dynamo}.
Here, we aim to provide a first answer to this question.

Double-diffusive fluids display a rich range of behaviours, with very different turbulent transport properties to single-component convection \citep[e.g.][]{garaud2018double}. One notable feature is the tendency to form density layers \citep[e.g.][]{pruzina2025one}, where convective wide regions of relatively uniform density are separated by sharp interfaces. This layering process is so ubiquitous that in the literature, semiconvection is often referred to simply as `layered convection'. It is well-documented in simulations that over time, a stack of several layers gradually coarsens via layer merger events, eventually leaving a single layer/interface in the domain \cite[e.g.][]{radko2007mechanics,pruzina2025one}. Much of the recent work on rotating semiconvection has focussed on this multiple-layered regime \cite[e.g.][]{moll2016effect,fuentes2024evolution,fuentes20253d}, leaving the behaviour outside (and after) layered states yet to be fully explored.

Despite significant interest in planetary semiconvection in recent years, there have been few previous attempts to demonstrate a dynamo maintained by semiconvection. \citet{mather2021regimes} concluded that it was only possibly for unrealistically large values of the magnetic Prandtl number, which characterises the ratio of viscous to magnetic diffusion, and thus semiconvection dynamos were unlikely to be relevant to stellar and planetary magnetic fields. However, their simulations did not show evidence of layering, and thus had very weak flow strengths, which may explain the inefficiency of their dynamo. By exploring a wider range of parameter space, we obtain a self-sustaining dynamo driven by convective motions in a layered state, for parameter values relevant to Jupiter, with a realistic dipolar form, and ratio of magnetic to kinetic energies. 

While our primary motivation is to explain the observed magnetic fields of planets in our solar system, dynamo action also supports stellar magnetic fields. \citep[e.g.][]{jones2010solar,charbonneau2014solar}. With semiconvective regions  also present in main sequence stars \citep[e.g.][]{merryfield1995hydrodynamics,spruit2013semiconvection,zaussinger2013semiconvection}, and likely to be present in exoplanets, the possibility of semiconvection to drive a dynamo has much broader scope, with the potential to significantly improve our understanding of astrophysical magnetic fields.

In this paper, we present simulation results demonstrating a semiconvection dynamo in a rapidly rotating spherical shell, for low magnetic Prandtl numbers. In Sect.~\ref{sec:equations}, we present the physical setup and governing equations for the problem. In Sect.~\ref{sec:jupiterparameters} we discuss the parameter ranges and expected structure of a Jupiter-like magnetic field; and in Sect.~\ref{sec:simulations} we present simulation results and compare them with the predictions of Sect.~\ref{sec:jupiterparameters}. Finally, in Sect.~\ref{sec:discussion} we summarise our findings and discuss the potential for future studies.

\section{Governing equations}\label{sec:equations}
\begin{figure}
    \centering
     \includegraphics[width=8cm]{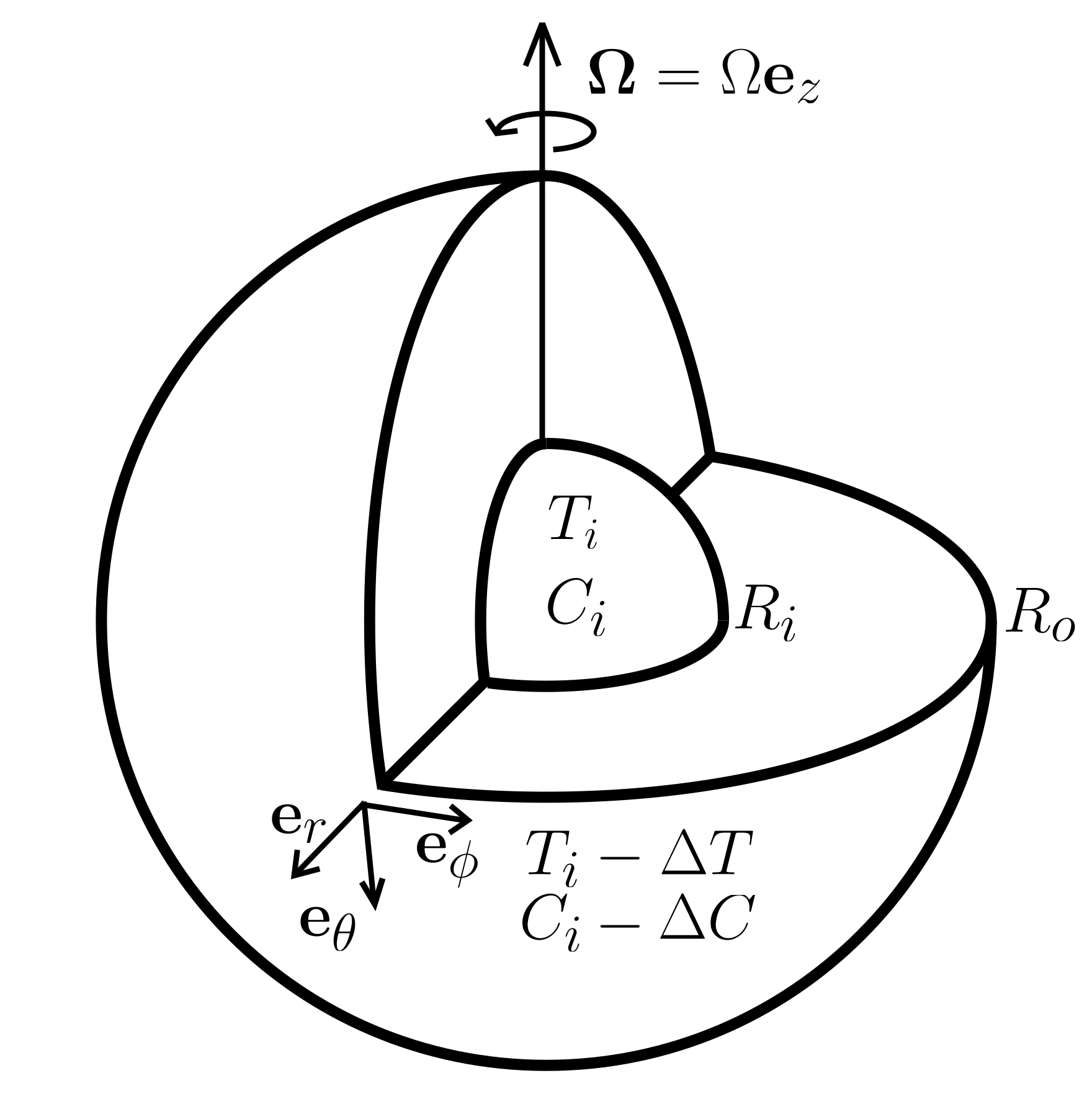}
    \caption{Outline of the physical setup of the problem.}
    \label{fig:setupsketch}
\end{figure}

Based on the work of \citet{debras2019new}, we consider the semiconvection region to be a spherical shell $R_i\leq r\leq R_o$, rotating at rate $\Omega$ and sandwiched between two convective regions.  We assume that the temperature and composition of heavy elements in these convective regions are fixed, with $T(R_i)-T(R_o)=\Delta T$ and $C(R_i)-C(R_o)=\Delta C$.We adopt spherical polar coordinates, with radial, polar and azimuthal basis vectors $\{\vec{e}_r,\vec{e}_\theta,\vec{e}_\phi\}$ defined such $\theta=0$ when the position vector $\vec{r}$ is aligned with the rotation vector $\vec{\Omega}=\Omega \vec{e}_z$. A summary of this setup is shown in Figure~\ref{fig:setupsketch}.

We model the evolution  of the fluid velocity $\vec{u}=u_r\vec{e}_r+u_\theta\vec{e}_\theta+u_\phi\vec{e}_\phi$, reduced pressure $p$ (that includes centrifugal effects), temperature $T$, composition $C$ and magnetic field $\vec{b}$ with the Navier-Stokes equations in the presence of background temperature and composition fields $T_0(\vec{r})$ and $C_0(\vec{r})$, coupled with the magnetic induction equation. We adopt the Boussinesq approximation that the changes in the fluid density are small compared to the magnitude of the density itself. These equations are expressed in dimensionless form as 
\begin{eqnarray}
    \frac{\partial}{\partial t}\vec{u} + \left(\frac{2}{\Ek}\vec{e}_z+\nablab\times\vec{u}\right)\times\vec{u} = -\nablab p \nonumber\\\>+ (T-C)\vec{e}_r + (\nablab\times\vec{b})\times\vec{b} + \nabla^2\vec{u} ,\label{eqn:momentum}\\
    \frac{\partial}{\partial t}T + \vec{u}\cdot\nablab\left(T+T_0\right) = \frac{1}{\Pr}\nabla^2T,\label{eqn:temperature}\\
    \frac{\partial}{\partial t}C + \vec{u}\cdot\nablab\left(C+C_0\right) = \frac{1}{\Sc}\nabla^2C,\label{eqn:composition}\\
    \frac{\partial}{\partial t}\vec{b}= \nablab\times\left(\vec{u}\times\vec{b}-\frac{1}{\Pm}\nablab\times\vec{b}\right)\label{eqn:magnetic}\\
    \nablab\cdot\vec{u}=0,\\
    \nablab\cdot\vec{b}=0.\label{eqn:incompressibility}
\end{eqnarray}
We assume a constant gravitational acceleration, pointing in the radial direction $\vec{e}_r$.
These equations are nondimensionalised based on the outer sphere radius $R_o$ and viscous timescale $\tau_\nu=R_o^2/\nu$, where $\nu$ is the kinematic viscosity. The dimensional fields can be recovered as follows:
\begin{eqnarray}
    \vec{u}_{dim} = \frac{\nu}{R_o}\vec{u},&\quad&  \vec{b}_{dim}=\sqrt{\rho_0\mu_0}\frac{\nu}{R_o}\vec{b},\nonumber\\
    \quad T_{dim} = \frac{\nu^2}{g\alpha_TR_o^3}T,&\quad&  C_{dim} = \frac{\nu^2}{g\alpha_C R_o^3}C,
\end{eqnarray}
where $\alpha_T$ and $\alpha_C$ are the coefficients of thermal and compositional expansion, $g$ is gravitational acceleration$\rho_0$ is a reference density and $\mu_0$ the vacuum permeability.
The system \eqref{eqn:momentum}--\eqref{eqn:incompressibility} is  controlled by four dimensionless parameters: the Ekman number 
\begin{equation}
    \Ek=\frac{\nu}{R_o^2\Omega}\label{eqn:ekman}
\end{equation}
and the Prandtl, Schmidt and magnetic Prandtl numbers
\begin{equation}
    \Pr=\frac{\nu}{\kappa_T},\quad \Sc=\frac{\nu}{\kappa_C},\quad \Pm=\frac{\nu}{\eta},\label{eqn:prandtls}
\end{equation}
where $\kappa_T$, $\kappa_C$ and $\eta$ are the thermal, compositional and magnetic diffusivities. The ratio of the compositional to thermal diffusivities is called the Lewis number $L=\Sc/\Pr$. The shell thickness is nondimensionalised to  $1-\Delta R\leq r\leq 1$, where $\Delta R = (R_o-R_i)/R_o$. 

The background temperature and salinity fields are given by
\begin{equation}
    T_0=\frac{\Rat}{\Pr}\delta(r),\quad C_0=\frac{\Rac}{\Sc}\delta(r),\label{eqn:backgroundfields}
\end{equation}
where $\delta(r) = \left(1-\Delta R\right)\left(1-r\right)/r\Delta R$ is the spherically symmetric solution to Laplace's equation in a spherical shell. This fairly simple basic state is not meant to match the observations as the fluid state will evolve dynamically due to semiconvection. 
The thermal and compositional Rayleigh numbers are defined as
\begin{equation}
    \Rat = \frac{\alpha_Tg\Delta TR_o^3}{\kappa_T\nu},\quad \Rac = \frac{\alpha_Cg\Delta CR_o^3}{\kappa_C\nu},\label{eqn:rayleighs}
\end{equation}
where $\alpha_T$ and $\alpha_C$ are the thermal and compositional expansion coefficients, $g$ is the gravitational acceleration, and $\Delta T $ and $\Delta C$ are the dimensional thermal and compositional differences across the layer, as shown in figure~\ref{fig:setupsketch}. The background fields can be used to define the Brunt-V\"ais\"al\"a frequency $N$, representing the frequency of oscillations of a displaced particle in the stratified fluid:
\begin{equation}
    N^2 = -\frac{\Rat}{\Pr}+\frac{\Rac}{\Sc}.
\end{equation}
In a fluid that is statically stable to top-heavy convection, $N^2>0$. The semiconvection region is defined by a stabilising compositional gradient and destabilising temperature gradient, i.e. 
\begin{equation}
\Rat>0,\quad\Rac>0.
\end{equation}
Within this regime, the semiconvection instability will develop if
\begin{equation}
    1 \leq R_\rho^* = \left|\frac{\Rac/Sc}{\Rat/Pr}\right|\leq\frac{\Pr+1}{\Pr+\Pr/\Sc} ,
\end{equation}
where the quantity $R_\rho^*$ is referred to as the (inverse) density ratio \citep[e.g.][]{turner1979buoyancy}.

To approximate the convective regions above and below the shell, we adopt fixed temperature, fixed composition and stress-free velocity boundary conditions as follows.
\begin{equation}
\begin{rcases}
    T =0\\
    C =0\\
     \frac{\partial}{\partial r}\left(u_\theta/r\right)=0\\
     \frac{\partial}{\partial r}\left(u_\phi/r\right)=0\\
     \vec{e}_r\cdot\vec{u}=0\\
     \end{rcases}\text{ at } r = 1-\Delta R, 1.\label{eqn:boundaryconditions}
\end{equation}
 We assume that inner convective region is electrically conducting, as suggested by \citet{debras2019new}, with the same electrical conductivity than the semiconvection layer, and adopt an insulating boundary condition on the outer sphere, to represent the low conductivity of molecular hydrogen. At the inner boundary, $\vec{b}$ and $\partial\vec{b}/\partial r$ are set to be continuous. Indeed, we consider the same conductivity and permeability in the fluid and in the inner domain. We have also tested a setup with an insulating core; the results are not significantly different, and it is computationally more expensive. The total angular momentum is set to zero.

We use the XSHELLS code \citep{schaeffer2013efficient,schaeffer2017turbulent} to solve equations~\eqref{eqn:momentum}--\eqref{eqn:incompressibility}, in the presence of background fields \eqref{eqn:backgroundfields}, with boundary conditions \eqref{eqn:boundaryconditions}.

\section{Conditions for a Jupiter-like dynamo}\label{sec:jupiterparameters}
While the results that follow pertain to all gas giants (and are of general interest in magnetohydrodynamics), we wish to place our work in the context of observations. The Juno mission has provided our most detailed measurements so far, with a model of the Jovian magnetic field up to spherical harmonic degree $l=18$ \citep{connerney2022new}, allowing us to assess the significance of our results. As such, we adopt parameter values relevant for the semiconvection region of Jupiter. We discuss the implications of our work for Saturn in Sect.~\ref{sec:discussion}

Astrophysical fluids are characterised by small Prandtl number $\Pr\ll1$, large Lewis number $L\gg1$ \citep{garaud2018double}, as well as small magnetic Prandtl number $\Pm\ll1$  \citep{roberts2000geodynamo}. At leading order, Jupiter's radial magnetic field is strongly dipolar, with the dipolar fraction of the magnetic energy $f_{dip}\approx0.7$ \citep{connerney2022new}.

The huge scales of gaseous planets are still inaccessible for direct numerical simulations. For Jupiter, we estimate the thermal Rayleigh number and Ekman number, as defined by equations \eqref{eqn:rayleighs} and \eqref{eqn:ekman} (details given in Appendix \ref{app:dimlessparameters})
\begin{equation}
    \Rat=O(10^{36}), \quad \Ek=O(10^{-19}).
\end{equation}
To locate the correct parameter space for our simulations, we follow \citet{monville2019rotating}, who found that, for the fingering regime of double diffusive convection, the hydrodynamic behaviour could be characterised based on the stratification-rotation ratio
\begin{equation}
    \frac{N}{\Omega}=\Ek\sqrt{-\Rat/\Pr+\Rac/\Sc}\label{eqn:NOmega}
\end{equation}
For the semiconvection layer of Jupiter, 
\begin{equation}
0.01\lesssim N/\Omega \lesssim1.\label{eqn:NOmega_range}
\end{equation}
Details of these estimates are given in Appendix~\ref{app:dimlessparameters}.

From the kinetic energy, we define the Reynolds number (the ratio of inertial to viscous forces) as follows
\begin{equation}
    \Rey=\frac{\sqrt{2\langle E_u\rangle_V}R_o}{\nu},
\end{equation}
where $\langle\cdot\rangle_V$ is a space-average across the entire domain. We define the magnetic Reynolds number (the ratio of induction to magnetic diffusion) as :
\begin{equation}
    \Rm=\Pm\Rey,\label{eqn:Rem}
\end{equation}
For a dynamo to develop, induction must dominate diffusion, i.e. $\Rm\gg1$. The exact value depends on the geometry and flow dynamics, but $\Rm\geq O(10^3)$ is generally sufficient. It follows from equation~\eqref{eqn:Rem} that for a small-$\Pm$ dynamo, $\Rm$ must be large, requiring strong turbulent motions.

The flow structure of Jupiter takes the form of a series of zonal jets, alternating in direction east/west. Observations show a large number of these jets responsible for the striking striped pattern at the surface. At the surface, these are very strong, with characteristic velocities $U=O(10-100)~\unit{m.s^{-1}}$, however beneath the surface layer the velocities reduce sharply to $O(0.01-1)~\unit{m.s^{-1}}$ \citep{moore2019time}. \citet{christensen2024quenching} argue that this reduction in jet strength is due to interaction with the stably stratified semiconvection layer, where buoyancy and electromagnetic forces act as a brake on the flow.

The kinetic and magnetic energy densities can be calculated as
\begin{equation}
    e_U=\frac{1}{2}\rho U^2,\quad e_B = \frac{1}{2}\frac{B^2}{\mu_0},\label{eqn:energydensities}
\end{equation}
where $\rho$ is the fluid density, $\mu_0$ is the magnetic permeability of the fluid, $U$ and $B$ are the dimensional velocity and magnetic field strength. Taking values appropriate for the semiconvection layer $\rho = 10^3~\unit{kg.m^{-3}}$, $\mu_0=4\pi\times10^{-7}~\unit{kg.m.s^{-1}.A^{-2}}$ \citep{french2012ab}, $U=O(0.01-1)~\unit{m.s^{-1}}$ \citep{moore2019time} and $B=4\times10^{-4}~\unit{T}$ \citep{connerney2022new}, we estimate the energy ratio as 
\begin{equation}
    10^{-4}\lesssim\frac{e_B}{e_U}\lesssim 1.\label{eqn:energyratio_range}
\end{equation}
Hence, it is possible that the magnetic energy may be of a similar size to the kinetic energy, and as such the magnetic field may have significant effects on the flow structure via the Lorentz force.

\section{Simulation results}\label{sec:simulations}
In this section, we first briefly discuss the general behaviour of magnetohydrodynamic (MHD) simulations, and  the critical parameters for a dynamo to develop. We then describe in detail a single long simulation, chosen to demonstrate a self-sustaining dynamo in the planetary relevant regime of low $\Pm$. We consider the variation in the magnetic energy, the three-dimensional form of the fields, and the energy spectra, and show that these compare favourably to our expectations for Jupiter.

\subsection{Onset of dynamo action}
\begin{figure}[ht]
    \centering
    \includegraphics[width=\columnwidth]{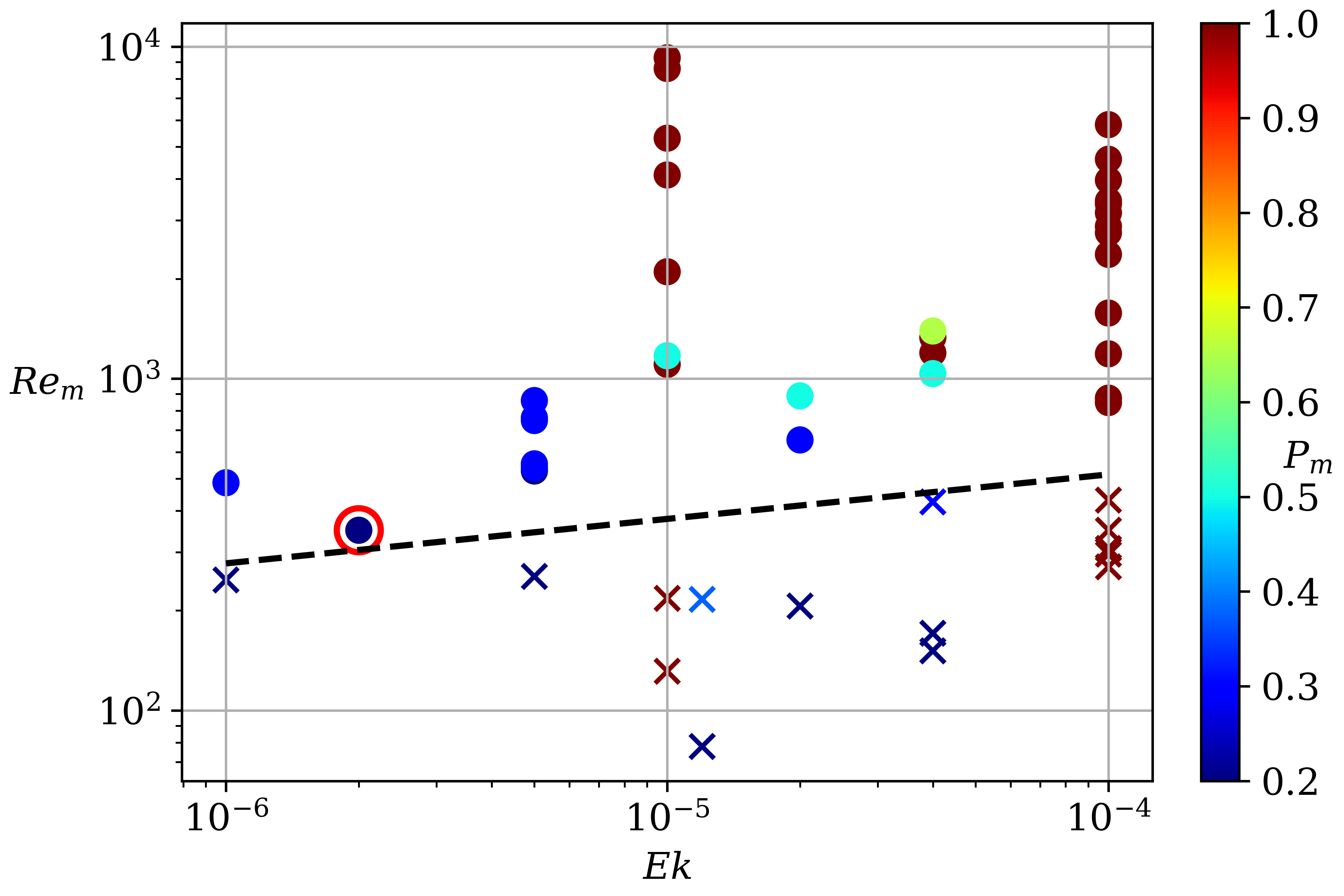}
    \caption{Stability of dynamo action in $\Ek$--$\Rm$ space, for a range of simulations, with varying values of $\Rac$, $\Rat$, $\Pm$ and $Ek$. Dots/crosses represent simulations showing growth/decay in the magnetic energy, coloured by the value of $\Pm$. The black dashed line is an estimate of the critical value $\Rm^C(\Ek)$. The red circle identifies the position of the simulation presented in detail in Sect.~\ref{sec:nonlinearsim}. The ranges of parameter values explored are given in Table~\ref{tab:simparams}.}
    \label{fig:onset}
\end{figure}

To find the suitable parameter ranges for dynamo action, we have conducted a number of simulations for a range of parameter values (listed in Table~\ref{tab:simparams}. With $\Pr=0.3$, $\Sc=3$ and $\Delta R=0.5$ fixed, we vary $\Ek$, $\Rac$, $\Rat$ and $\Pm$. We run hydrodynamic simulations until the semiconvection instability has saturated, then seed a random small-amplitude magnetic field. After an initial adjustment period, the simulation magnetic energy $E_b^S$ undergoes a linear growth or decay phase $E_b^S\sim\exp(\sigma t)$. Figure~\ref{fig:onset} shows a summary of the behaviour of simulations in $\Ek$--$\Rm$ space. For each value of $\Ek$, there is growth in the magnetic energy for $\Rm$ above a critical value $\Rm^C(\Ek)$, which appears to be roughly independent of $\Pm$. For $10^{-6}\leq\Ek\leq10^{-4}$, $250\lesssim\Rm^C\lesssim850$. Figure~\ref{fig:onset} shows that $\Rm^C$ increasing as $\Ek$ increases, meaning that simulations with relatively small values of $\Ek$ are favoured to achieve a dynamo at low $\Pm$.

\begin{table}[ht]
    \centering
    \caption{Summary of parameters for simulations in Fig.~\ref{fig:onset}}
    \begin{tabular}{cc}
        \hline\hline
         Parameter&Value(s) \\
         $\Rac$&$7.5\times10^7$ -- $8\times10^{10}$\\
         $\Rat$&$\Rac/12$\\
         $\Ek$&$2\times10^{-6}$ -- $1\times10^{-4}$\\
         $\Pm$&$0.2$--$1.0$\\
         $\Pr$&$0.3$\\
         $\Sc$&$3$\\
         $\Delta R$&$0.5$\\
         \hline
    \end{tabular}
    \label{tab:simparams}
\end{table}

\subsection{Nonlinear simulation of a Jupiter-like dynamo}\label{sec:nonlinearsim}
We show the results of a numerical simulation with $\Ek=2\times10^{-6}$, $\Rac=2\times10^{10}$, $\Rat=1.67\times10^9$ (so $R_\rho^*=1.2$), $\Pr=0.3$, $\Sc=3$, $\Pm=0.2$ and $\Delta R=0.5$. The spatial resolution is $384\times300\times300$ in $r$, $l$ and $m$ directions respectively, with $100$ of the radial shells reserved for the conducting inner sphere $0<r\leq0.5$. This simulation is marked on the $\Ek$--$\Rm$ plot in Fig.~\ref{fig:onset} with a red circle. With these values, the stratification-rotation ratio is $N/\Omega=0.067$, at the low end of our estimate \eqref{eqn:NOmega_range}. Notably, the magnetic Prandtl number is significantly less than unity, in contrast to the work of \citet{mather2021regimes}. 

\begin{figure}
    \centering
    \includegraphics[width=1\columnwidth]{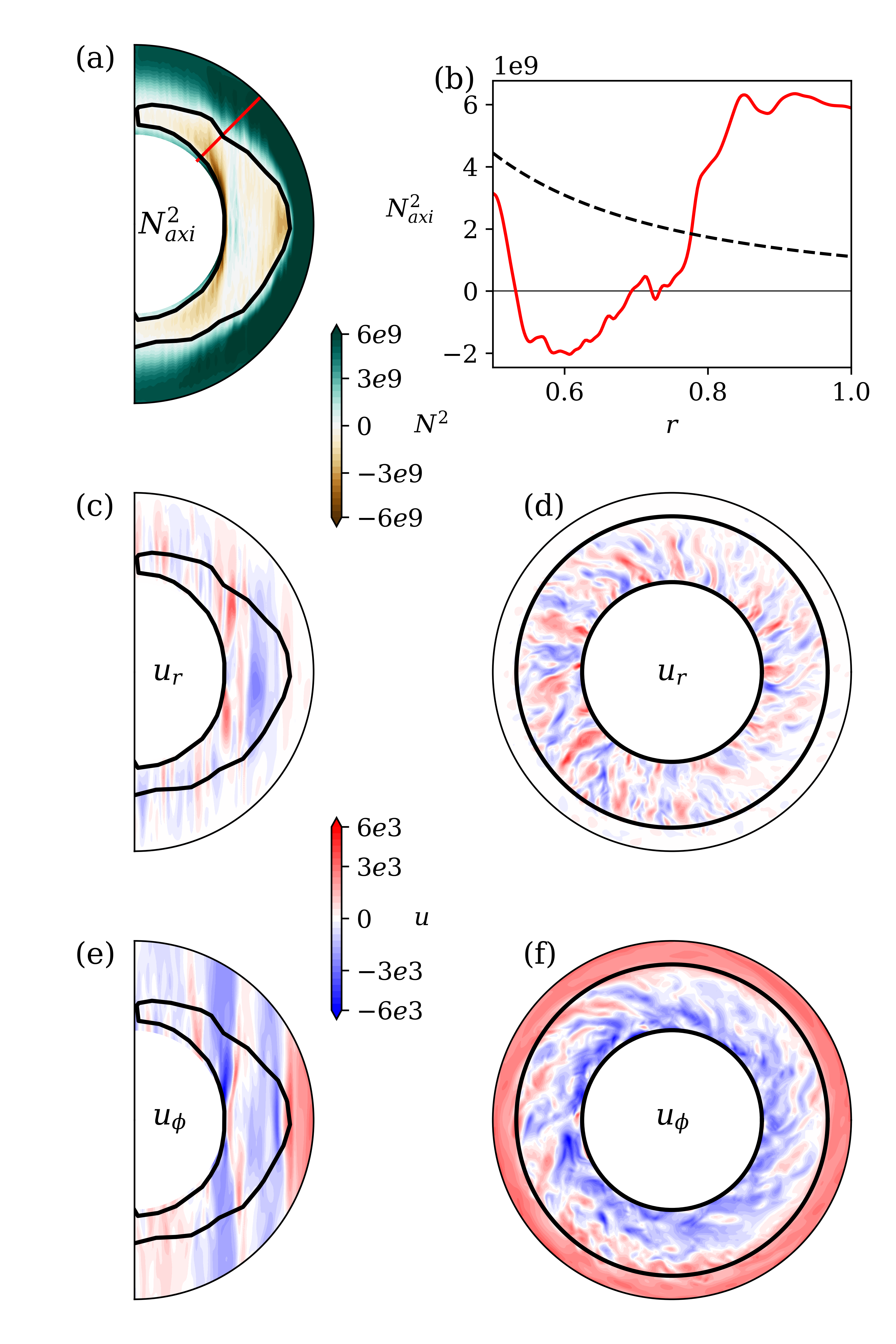}
    \caption{Initial condition for the main simulation, showing the axisymmetric part $N^2_\text{axi}$ of the square of the Brunt V\"ais\"al\"a frequency (a) in a meridional slice and (b) along the transect shown in red in (a). The black dashed line shows the background conductive state. (c)--(f) Meridional and equatorial slices of the radial and azimuthal velocity fields $u_r$ and $u_\phi$. Solid black lines on the snapshots show the boundaries of the convective region, where $N^2_{axi}<0$.}
    \label{fig:initial_fields}
\end{figure}

The initial conditions are shown in Fig.~\ref{fig:initial_fields}. These conditions represent the long-term statistically stable state of a hydrodynamic ($\bm{b}=\bm{0}$) simulation. From an initially stable stratification, Fig.~\ref{fig:initial_fields}(a)--(b) show that a convective layer has spontaneously developed in the inner part of the shell where the  Brunt-V\"ais\"al\"a frequency is negative. Outside this region, the fluid remains stably stratified. This represents the final state of layered convection, after all other layers have merged. The snapshots in Fig.~\ref{fig:initial_fields}(c)--(d) show that the radial velocities are significantly larger in the convective region than in the stably stratified layer. The meridional velocity field in Fig.~\ref{fig:initial_fields}(e)--(f) shows a zonal jet structure, with a strong prograde equatorial jet, and generally (but not exclusively) retrograde flow  at higher latitudes, extending in columns through the interior of the domain.

\begin{figure}
    \centering
    \includegraphics[width=\columnwidth]{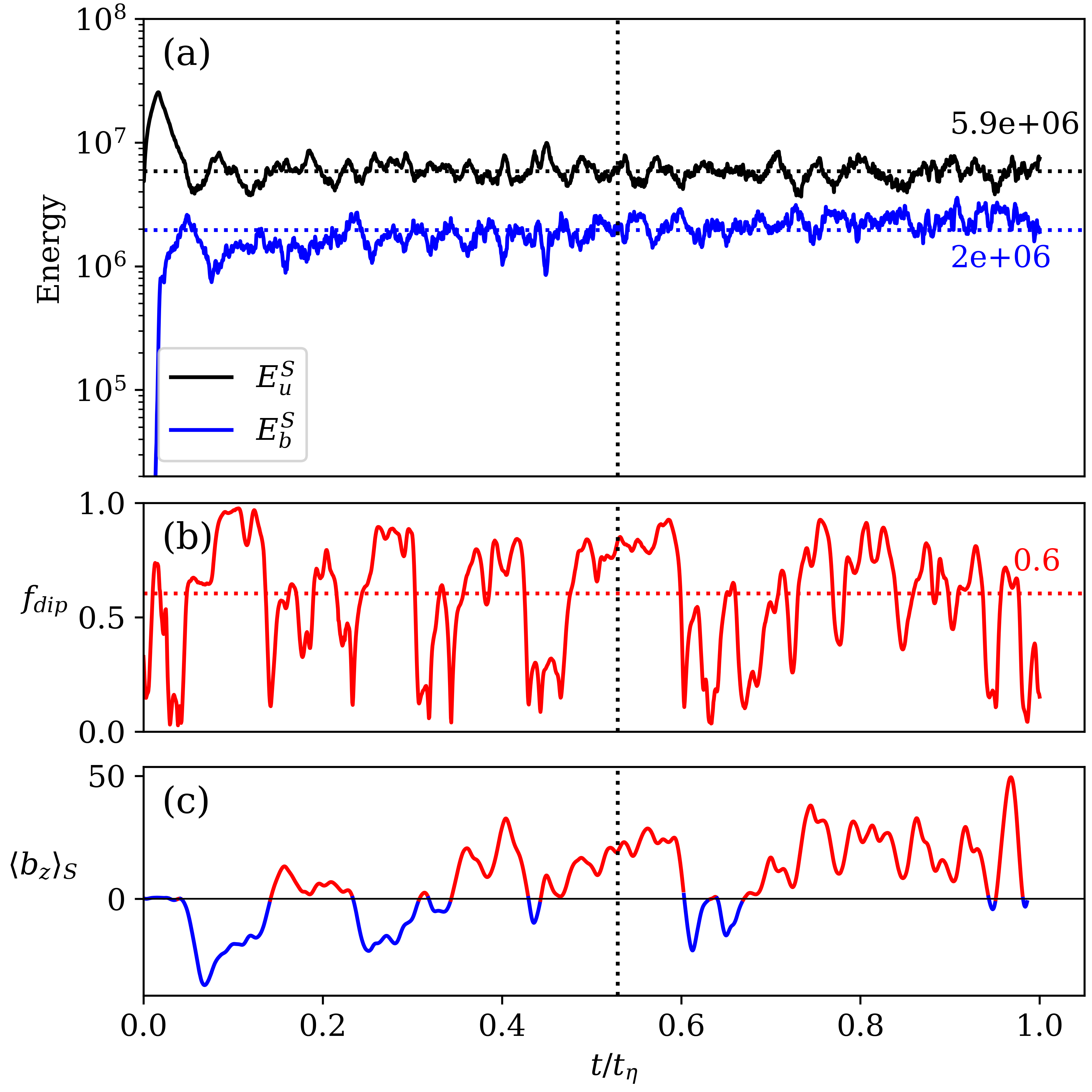}
    \caption{Time-evolution of (a) the kinetic and magnetic energies $E_u^S$ and $E_b^S$, (b) the dipolar energy fraction $f_{dip}$, and (c) the mean radial magnetic field on the outer sphere $\langle b_r\rangle_S$, for a MHD simulation with $\Ek=2\times10^{-6}$, $\Rac=2\times10^{10}$, $\Rat=1.67\times10^9$, $\Pr=0.3$, $\Sc=3$, $\Pm=0.2$ and $\Delta R=0.5$. Mean values are marked with horizontal dotted lines; The vertical dotted line at $t=0.53t_\eta$ shows the time at which snapshots are shown.}
    \label{fig:energies}
\end{figure}

Figure~\ref{fig:energies} shows the evolution of the kinetic and magnetic energies $E_u^S$ and $E_b^S$ in the simulation, and dipolar energy fraction $f_{dip}$, over a full magnetic diffusion time $t_\eta=R_o^2/\eta$. After an initial adjustment period, both the kinetic and magnetic energies remain relatively constant aside from some high frequency oscillations, with $E_b^S/E_u^S\approx0.3\pm0.2$. This ratio fits well within the expected range \eqref{eqn:energyratio_range}. The magnetic Reynolds number is approximately $\Rm\approx 350$; very close to the onset of magnetic instability, as seen in Fig.~\ref{fig:onset}. The dipole fraction $f_{dip}$ shows some variation over time, but has mean $0.6$, and takes a value $f_{dip}>0.5$ at approximately $70\%$ of times. These values are all consistent with those expected in Jupiter, as discussed in Sect.~\ref{sec:jupiterparameters}. 

\begin{figure*}
    \centering
    \includegraphics[width=0.8\textwidth]{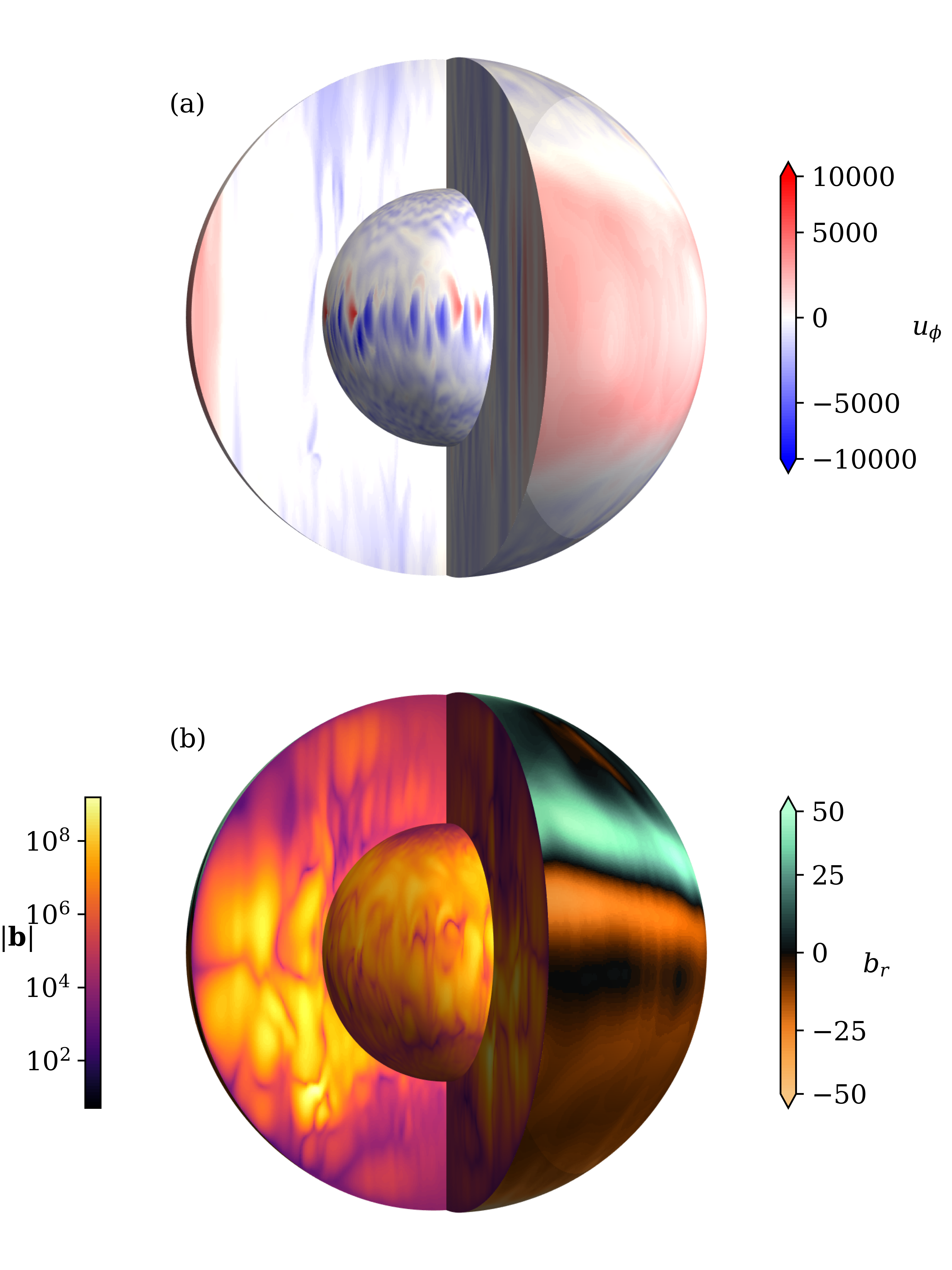}
    \caption{(a) Azimuthal velocity field shown on both spherical surfaces and meridional cross sections at $\phi=0$ and $3\pi/2$ at time $t=0.53t_\eta$. The flow takes the form of a strong prograde equatorial surface jet, with retrograde motion near the poles and in the interior. (b) Spatial structure of the magnetic field, with $b_r$ shown on the outer boundary, and $|\vec{b}|$ in the interior. The radial field at the surface is strongly dipolar, and the interior field shows very localised activity. The surface field is much weaker than in the interior. }
    \label{fig:spherecuts}
\end{figure*}

We now consider the 3d form of the flow and the magnetic field. Figure~\ref{fig:spherecuts}(a) shows a snapshot of the zonal flow $u_\phi$. There is a strong prograde zonal jet near the equator, with retrograde flow nearer the poles. The flow is structured in columns aligned with the rotation axis, extending through the entire fluid depth.

Figure~\ref{fig:spherecuts}(b) shows the magnetic field in a dipolar phase, with $|\vec{b}|$ plotted in the interior and $b_r$ shown on the surface $r=R_o$. On the outer boundary, the field is strongly dipolar. The interior field consists of a comparatively weak background field, with localised patches with much greater magnitudes. These hot spots appear to align with the region between the prograde jet and the tangent cylinder to the inner sphere, suggesting that they may be generated by the strong shear in this area, similarly to the work of \citet{guervilly2012dynamo}.

\begin{figure}
    \centering
    \includegraphics[width=\columnwidth]{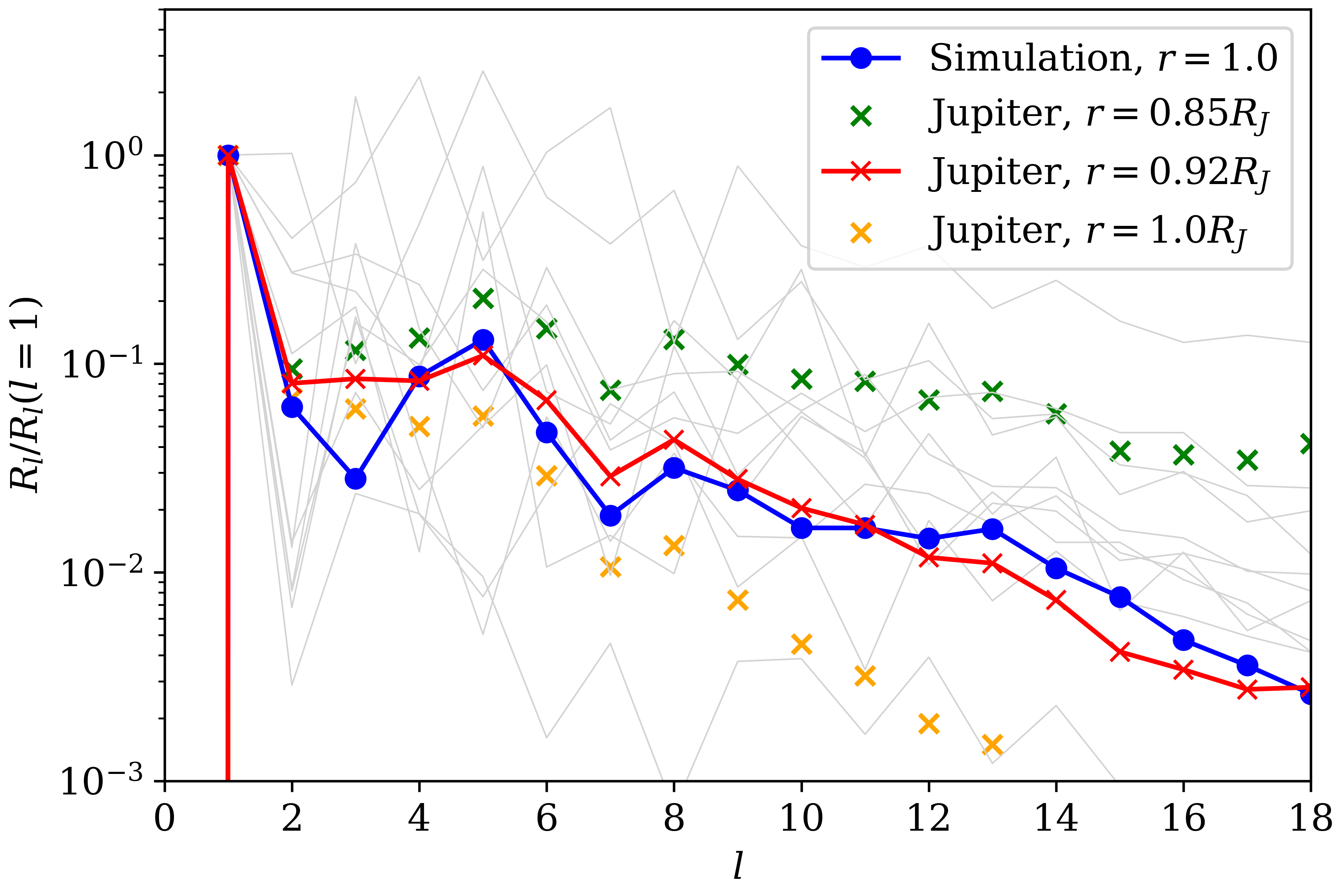}
    \caption{Lowes (magnetic energy) spectrum, normalised by the $l=1$ term, for (blue) the simulation data taken at $r=1$ at time $0.53t_\eta$, and  Juno observations of Jupiter \citep{connerney2022new}, taken at (green) $r=0.85R_J$, (red) $r=0.92R_J$ and (orange) $r=1.0R_J$. Grey lines show the simulation energy spectrum at a range of other times.}
    \label{fig:spectra}
\end{figure}

To assess how closely our simulations reproduce the magnetic field of Jupiter, we compare the Lowes spectrum (magnetic energy spectrum at the surface) \citep{lowes1974spatial} with the most recent observational model \citep{connerney2022new}, provided by the \texttt{planetMagFields} python library \citep{Barik2024}. As our simulations are intended to model the semiconvection layer beneath the surface, we must compare the simulation data at $r=1$ with the observational data some distance beneath the planetary surface. Fig.~\ref{fig:spectra} shows the Lowes spectrum, normalised by the $l=1$ component, for the simulation at $t=0.53t_\eta$ in blue, with a range of other times in light grey. and observational data at three radii: $0.85$, $0.92$ and $1.0R_J$. The simulation data at the highlighted time agrees well with the intermediate Jupiter spectrum ($r=0.92R_J$), with the interior spectrum being flatter, and the surface spectrum being steeper, while the other spectra show variation between rather dipolar profiles that correspond well with the highlighted profile, and some non-dipolar profiles with flatter spectra. Note that at the surface of the simulation, the magnetic field spectrum is rarely flat, contrasting with the usual assumption that the top of the dynamo region can be identified by a flat spectrum \citep[e.g.][]{connerney2022new}.

To compare the magnetic field strength in our simulations to that of Jupiter, we rescale the field based on the ratio of the simulation kinetic energy $E_u^S$ to estimated true kinetic energy density in the semiconvection zone of Jupiter $e_U$. Appendix \ref{app:dimfieldstrength} details the calculation, giving a scaling factor $Q=e_U/E_u^S=8.33\times10^{-5}\unit{kg m^{-1}s^{-2}}$. To obtain an estimate of the redimensionalised magnetic field from the simulation, the dimensionless $b$ must be multiplied by the dimensional factor $\sqrt{\mu_0Q}=0.014 \unit{T}$.

\begin{figure}
    \centering
    \includegraphics[width=\columnwidth]{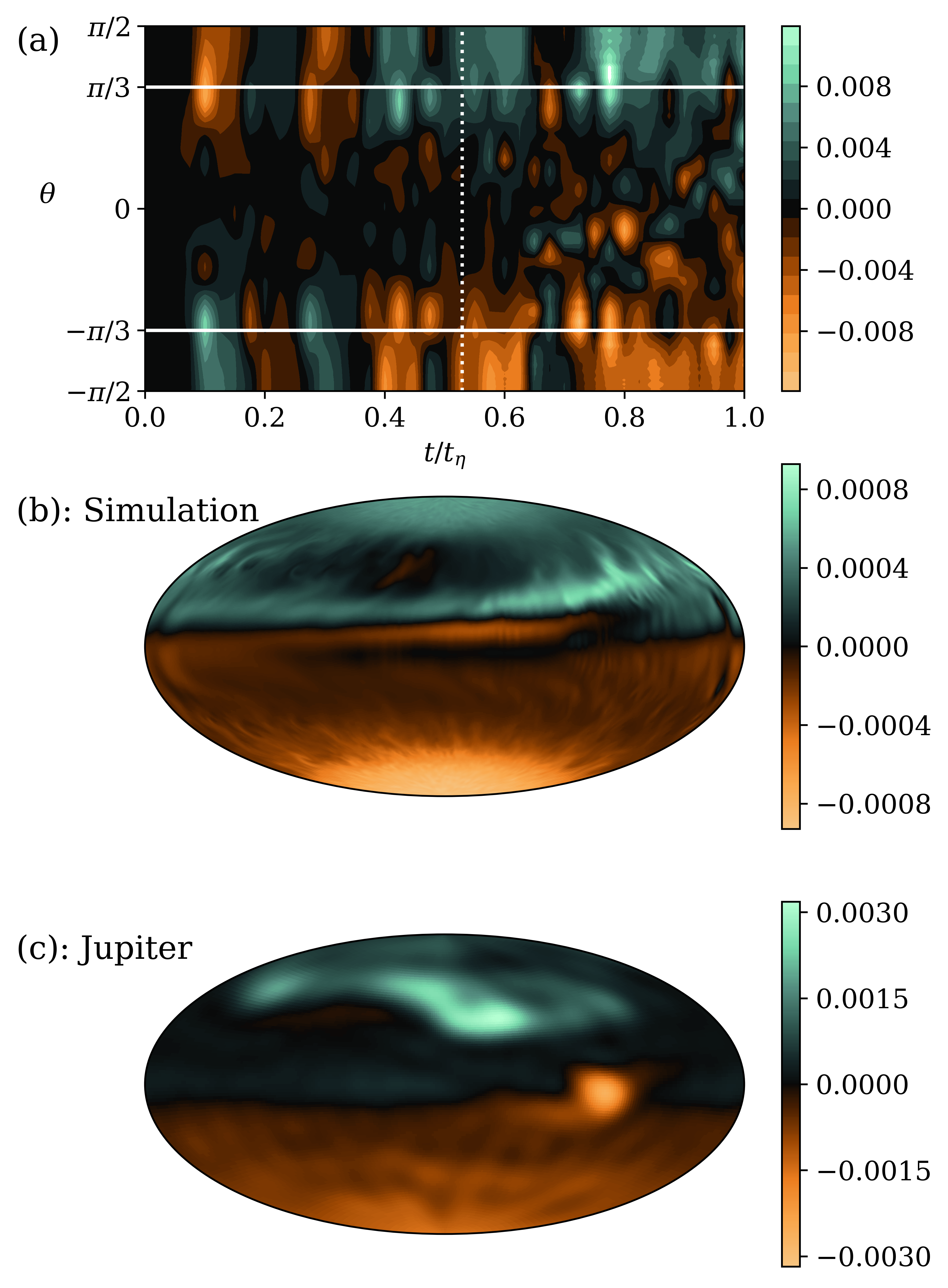}
    \caption{(a) Time-evolution of the axisymmetric (m=0) component of the dimensionalised surface radial field in Tesla, $\tilde{B}_r=\sqrt{\mu_0Q}b_r$. Solid white lines show the tangent cylinder to the inner sphere; the dotted white line shows the time at which the snapshot is taken. (b) Surface radial magnetic field $\tilde{B}_r$ in Tesla for the simulation at time $0.53t_\eta$. (c) Radial magnetic field of Jupiter  at $r=0.92R_J$, from \citep{Barik2024}.}
    \label{fig:surffield}
\end{figure}

Fig.~\ref{fig:surffield}(a) shows how the estimated redimensionalised surface magnetic field varies in time. The overall dipolar structure is clearly visible, with some short-timescale smaller scale structures visible near the equator, outside the tangent cylinder of the inner sphere. The field undergoes a number of reversals, associated with times when the dipole becomes less dominant. 

Figs.~\ref{fig:surffield}(b) and (c) show, respectively, the rescaled surface radial magnetic field $\sqrt{\mu_0Q}b_r$ in the simulation at time $0.53t_\eta$, and the radial magnetic field of Jupiter at $r=0.92R_J$. Both plots show an overall dipolar form, with some higher $l$ structures near the equator, and a similar magnetic field strength ($\sim 0.8~\unit{mT}$ for the simulation, $\sim 3~\unit{mT}$ for the observational data). The simulation data is much more regular than the observational data, with the field in Jupiter exhibiting the well-known large flux patch in the northern hemisphere, and strong negative patch near the equator \citep{connerney2022new}

\section{Discussion \& conclusions}\label{sec:discussion}
In this paper, we have demonstrated the possibility of semiconvection driving a planetary dynamo at low magnetic Prandtl number. Thanks to the Juno mission, there is plentiful data from Jupiter, which we have aimed our study to recreate. Our simulation results show striking resemblance to the Jovian magnetic field, with a strongly dipolar surface field for the majority of the time, and higher-$l$ structures visible near the equator. The ratio of magnetic to kinetic energies lies within our predicted range, and an estimation based on this range obtains a very similar magnetic field strength to observational data. The existence of a semiconvection dynamo at low $\Pm$ runs contrary to the work of \citet{mather2021regimes}, who concluded that unrealistically high values of $\Pm$ would be required, and thus semiconvection was unlikely to be important for planetary magnetic fields. The simulations of \citet{mather2021regimes} were performed at a relatively slow rotation rate $Ek\approx10^{-4}$ and relatively modest temperature and composition gradients $\Rat=O(10^5)$, $\Rac=O(10^6)$. By contrast, we take $\Ek=2\times10^{-6}$, $\Rac=2\times10^{10}$ and $\Rat=1.67\times10^9$. These values, while still far from realistic astrophysical parameters, result in a semiconvection field with much higher $\Rey$, and therefore a value of $\Rm$ above the dynamo threshold can be achieved for much smaller $\Pm$.

In our simulation, the hydrodynamic layering instability acting on a basic semiconvection field produced an inner convective region underneath a stably stratified region, with convection in this inner region driving dynamo action. This demonstrates a mechanism for magnetic field generation in stably stratified regions of planets, not by the primary semiconvection instability \cite[as investigated by][]{mather2021regimes}, but by the secondary layered state. 

Here, we consider parameter values that are fairly modest compared to true planetary values, but nonetheless provide computational challenges. Simulations with these values are made possible by the tremendous efficiency of the XSHELLS solver \citep{schaeffer2013efficient, monville2019rotating}, and in particular its GPU implementation. Thus, any more extreme parameters would require incredible large numbers of GPU-hours. A key difference from a more realistic planetary setup is the aspect ratio; \citet{debras2019new} suggest a maximum thickness of $0.25R_J$ for the semiconvection layer --- much less than our $\Delta R = 0.5$. However the reduction in layer depth proves to be rather costly computationally, with a significantly larger spectral resolution for similar parameters. At $\Pm=0.2$, we are also near the maximum value that could truly be considered `low-$\Pm$'. An extension to even lower values would be an useful development, as would an investigation of how the form of the magnetic field changes as $\Pm$ is varied.

We have adopted the Boussinesq approximation, which neglects density differences unless they appear in terms multiplied by gravity. This is essentially the assumption that the stably stratified layer is sufficiently small that the density does not vary significantly across it. For accurate planetary applications, the anelastic approximation is potentially more appropriate \citep[e.g.][]{yadav2013consistent}. However, anelastic models are more computationally costly, and may prevent the exploration of such extreme parameter values. Moreover, it is unclear how the well-mixed assumption required by anelastic models is compatible with the presence of stably stratified layers. With \citet{yadav2013consistent} showing that scaling laws of Boussinesq and anelastic dynamos are essentially the same, the Boussinesq equations remain thus an appropriate choice for an initial investigation into the possibility of semiconvection dynamos.

In this paper, we have focused on the magnetic field of Jupiter due to the quality of the data from the Juno mission. However, the Cassini mission has also obtained measurements of Saturn's magnetic field up to degree $l=11$ \citep{dougherty2018saturn}. The field is notably more axisymmetric than Jupiter's, with a dominant dipole component, and remarkably little variation in the azimuthal variation, even when the large-scale field is neglected. Compared to Jupiter, both the dynamo radius, and height of the stably stratified region are thought to be significantly deeper \citep{cao2012saturn}. The existence of a thin stably stratified semiconvection layer above a deep convective dynamo has been suggested as an explanation for the axisymmetry of the magnetic field, with non-axisymmetric components attenuated by zonal flows in the stable layer \citep{fuller2014saturn,stanley2016secular}. However other models have proposed that the stably-stratified region extends deep into the core \citep{mankovich2021diffuse}. If a stably stratified layer can both generate a magnetic field through semiconvection, and filter out the non-axisymmetric components, this could potentially resolve these contradictory proposals. (See, for example, Fig.~\ref{fig:spherecuts}, where the magnetic field is generated relatively deeply, whereas the strongest zonal flows are on the surface of the domain.)
In addition, some of our preliminary simulations operate in a non-dipolar regime; future work could explore the possibility of these to be relevant for the magnetic fields of Uranus and Neptune.

Of course, this is only a preliminary demonstration of a subject that is sure to provoke significant further research. The semiconvective instability in a rotating spherical geometry is still poorly understood, and a thorough study of both the onset of semiconvection and the form of the resultant flows is needed. A more detailed investigation of the dynamo, considering a wide range of parameter space, would allow general trends to be identified, and the results to be extrapolated to true planetary scales. We hope to report on both of these themes in the near future.

\begin{acknowledgements}
DC thanks F. Debras for discussions about the possibility of dynamo effects within semiconvection layers of gaseous planets. Work funded by the ERC under the European Union’s Horizon 2020 research and innovation program via the THEIA project (grant agreement no. 847433). Computations on the GRICAD infrastructure (https://gricad.univ-grenoble-alpes.fr), and HPC resources (Jean Zay V100 and H100) of IDRIS under allocation AD010413621 and A0160407382 attributed by GENCI (Grand Equipement National de Calcul Intensif). \textit{Author contribution statement.} PP produced the simulations, the figures, and the draft. DC came up with the initial research concept, and obtained the funding. NS modified the numerical code and helped for computations. DC and NS obtained computation hours from GENCI. All authors discussed each step of the work and the results; all edited the final manuscript.
\end{acknowledgements}

\bibliography{bibliography}

\begin{thebibliography}{40}
\expandafter\ifx\csname natexlab\endcsname\relax\def\natexlab#1{#1}\fi

\bibitem[{Barik \& Angappan(2024)}]{Barik2024}
Barik, A. \& Angappan, R. 2024, JOSS, 9, 6677

\bibitem[{Cao {et~al.}(2012)Cao, Russell, Wicht, Christensen, \& Dougherty}]{cao2012saturn}
Cao, H., Russell, C.~T., Wicht, J., Christensen, U.~R., \& Dougherty, M.~K. 2012, Icarus, 221, 388

\bibitem[{Charbonneau(2014)}]{charbonneau2014solar}
Charbonneau, P. 2014, AAR\&A, 52, 251

\bibitem[{Christensen \& Wulff(2024)}]{christensen2024quenching}
Christensen, U.~R. \& Wulff, P.~N. 2024, PNAS, 121, e2402859121

\bibitem[{Connerney {et~al.}(2022)Connerney, Timmins, Oliversen, Espley, Joergensen, Kotsiaros, Joergensen, Merayo, Herceg, Bloxham, {et~al.}}]{connerney2022new}
Connerney, J., Timmins, S., Oliversen, R., {et~al.} 2022, JGRE, 127, e2021JE007055

\bibitem[{Debras \& Chabrier(2019)}]{debras2019new}
Debras, F. \& Chabrier, G. 2019, ApJ, 872, 100

\bibitem[{Dougherty {et~al.}(2018)Dougherty, Cao, Khurana, Hunt, Provan, Kellock, Burton, Burk, Bunce, Cowley, {et~al.}}]{dougherty2018saturn}
Dougherty, M.~K., Cao, H., Khurana, K.~K., {et~al.} 2018, Science, 362, eaat5434

\bibitem[{Duarte {et~al.}(2018)Duarte, Wicht, \& Gastine}]{duarte2018physical}
Duarte, L.~D., Wicht, J., \& Gastine, T. 2018, Icar, 299, 206

\bibitem[{French {et~al.}(2012)French, Becker, Lorenzen, Nettelmann, Bethkenhagen, Wicht, \& Redmer}]{french2012ab}
French, M., Becker, A., Lorenzen, W., {et~al.} 2012, ApJS, 202, 5

\bibitem[{Fuentes(2025)}]{fuentes20253d}
Fuentes, J. 2025, ApJ, 982, 44

\bibitem[{Fuentes {et~al.}(2024)Fuentes, Hindman, Fraser, \& Anders}]{fuentes2024evolution}
Fuentes, J., Hindman, B.~W., Fraser, A.~E., \& Anders, E.~H. 2024, ApJL, 975, L1

\bibitem[{Fuller(2014)}]{fuller2014saturn}
Fuller, J. 2014, Icarus, 242, 283

\bibitem[{Garaud(2018)}]{garaud2018double}
Garaud, P. 2018, AnRFM, 50, 275

\bibitem[{Gastine {et~al.}(2012)Gastine, Duarte, \& Wicht}]{gastine2012dipolar}
Gastine, T., Duarte, L., \& Wicht, J. 2012, A\&A, 546, A19

\bibitem[{Guervilly {et~al.}(2012)Guervilly, Cardin, \& Schaeffer}]{guervilly2012dynamo}
Guervilly, C., Cardin, P., \& Schaeffer, N. 2012, Icar, 218, 100

\bibitem[{Jones(2014)}]{jones2014dynamo}
Jones, C. 2014, Icar, 241, 148

\bibitem[{Jones(2011)}]{jones2011planetary}
Jones, C.~A. 2011, AnRFM, 43, 583

\bibitem[{Jones {et~al.}(2010)Jones, Thompson, \& Tobias}]{jones2010solar}
Jones, C.~A., Thompson, M.~J., \& Tobias, S.~M. 2010, SSRv, 152, 591

\bibitem[{Leconte \& Chabrier(2012)}]{leconte2012new}
Leconte, J. \& Chabrier, G. 2012, A\&A, 540, A20

\bibitem[{Leconte \& Chabrier(2013)}]{leconte2013layered}
Leconte, J. \& Chabrier, G. 2013, Nat. Geosci., 6, 347

\bibitem[{Lowes(1974)}]{lowes1974spatial}
Lowes, F. 1974, GeoJI, 36, 717

\bibitem[{Mankovich \& Fuller(2021)}]{mankovich2021diffuse}
Mankovich, C.~R. \& Fuller, J. 2021, Nat. Astron., 5, 1103

\bibitem[{Mather \& Simitev(2021)}]{mather2021regimes}
Mather, J.~F. \& Simitev, R.~D. 2021, GApFD, 115, 61

\bibitem[{Merryfield(1995)}]{merryfield1995hydrodynamics}
Merryfield, W.~J. 1995, ApJ, 444, 318

\bibitem[{Moll \& Garaud(2016)}]{moll2016effect}
Moll, R. \& Garaud, P. 2016, ApJ, 834, 44

\bibitem[{Monville {et~al.}(2019)Monville, Vidal, C{\'e}bron, \& Schaeffer}]{monville2019rotating}
Monville, R., Vidal, J., C{\'e}bron, D., \& Schaeffer, N. 2019, GeoJI, 219, S195

\bibitem[{Moore {et~al.}(2022)Moore, Barik, Stanley, Stevenson, Nettelmann, Helled, Guillot, Militzer, \& Bolton}]{moore2022dynamo}
Moore, K., Barik, A., Stanley, S., {et~al.} 2022, JGRE, 127, e2022JE007479

\bibitem[{Moore {et~al.}(2019)Moore, Cao, Bloxham, Stevenson, Connerney, \& Bolton}]{moore2019time}
Moore, K., Cao, H., Bloxham, J., {et~al.} 2019, NatAs, 3, 730

\bibitem[{Pru{\v{z}}ina(2025)}]{pruzina2025one}
Pru{\v{z}}ina, P. 2025, JFM, 1008, A29

\bibitem[{Radko(2007)}]{radko2007mechanics}
Radko, T. 2007, JFM, 577, 251

\bibitem[{Radko(2013)}]{radko2013double}
Radko, T. 2013, Double-diffusive convection (Cambridge University Press)

\bibitem[{Roberts \& Glatzmaier(2000)}]{roberts2000geodynamo}
Roberts, P.~H. \& Glatzmaier, G.~A. 2000, RvMP, 72, 1081

\bibitem[{Schaeffer(2013)}]{schaeffer2013efficient}
Schaeffer, N. 2013, GGG, 14, 751

\bibitem[{Schaeffer {et~al.}(2017)Schaeffer, Jault, Nataf, \& Fournier}]{schaeffer2017turbulent}
Schaeffer, N., Jault, D., Nataf, H.-C., \& Fournier, A. 2017, GeoJI, 211, 1

\bibitem[{Schwarzschild \& H{\"a}rm(1958)}]{schwarzschild1958evolution}
Schwarzschild, M. \& H{\"a}rm, R. 1958, ApJ, vol. 128, p. 348, 128, 348

\bibitem[{Spruit(2013)}]{spruit2013semiconvection}
Spruit, H. 2013, A\&A, 552, A76

\bibitem[{Stanley \& Bloxham(2016)}]{stanley2016secular}
Stanley, S. \& Bloxham, J. 2016, Phys. Earth and Planet. Inter., 250, 31

\bibitem[{Turner(1979)}]{turner1979buoyancy}
Turner, J.~S. 1979, Buoyancy effects in fluids (Cambridge University Press)

\bibitem[{Yadav {et~al.}(2013)Yadav, Gastine, Christensen, \& Duarte}]{yadav2013consistent}
Yadav, R.~K., Gastine, T., Christensen, U.~R., \& Duarte, L.~D. 2013, ApJ, 774, 6

\bibitem[{Zaussinger \& Spruit(2013)}]{zaussinger2013semiconvection}
Zaussinger, F. \& Spruit, H. 2013, A\&A, 554, A119

\end{thebibliography}
\bibliographystyle{aa}

\begin{appendix}
\section{Dimensionless parameters}\label{app:dimlessparameters}
To calculate the values of $\Rat$, $\Ek$ and $N/\Omega$ given in Sect.~\ref{sec:jupiterparameters}, we summarise the values of key physical parameters in Table~\ref{tab:parameters}. We choose the radii $0.65R_J$ and $0.95R_J$ for the temperature to give an intermediate size for the stratified region; the exact points chosen will have only a small effect on the order of magnitude estimates. 

According to the definitions in equations \eqref{eqn:ekman}-\eqref{eqn:prandtls} and \eqref{eqn:rayleighs}, from the data in Table~\ref{tab:parameters}, we obtain the following dimensionless quantites
\begin{eqnarray}
    \Pr\approx0.072,\quad \Sc\approx2,\quad\Pm\approx1.6\times10^{-7},\nonumber\\\>\quad\Rat\approx-1.1\times10^{36},\quad\Ek\approx4.2\times10^{-19}. 
\end{eqnarray}
\citet{debras2019new} gives the density ratio in the semiconvection layer to be in the range $1\leq R_\rho^*\leq3.3$. For this range, the stratification-rotation ratio is given by equation~\eqref{eqn:NOmega} to be 
\begin{equation}
    0\leq \frac{N}{\Omega}\lesssim2.5,
\end{equation}
with a value of $N/\Omega\approx0.7$ at $R_\rho^*=1.2$.
\begin{table}[ht]
    \centering
    \caption{Values of physical quantities required for the estimations.}
    \begin{tabular}{ccc}
        \hline\hline
         Quantity&Value&Units \\
         \hline
         $\alpha_T$&$3.6\times10^5$&$\unit{K^{-1}}$\\
         $\kappa_T$&$5\times10^{-6}$&$\unit{m^2.s^{-1}}$\\
         $\kappa_C$& $0.18\times10^{-6}$&$\unit{m^2.s^{-1}}$\\
         $\eta$&$2.21$&$\unit{m^2.s^{-1}}$\\
         $\nu$&$0.36\times10^{-6}$&$\unit{m^2.s^{-1}}$\\
         $\rho$&$820$&$\unit{kg.m^{-3}}$\\
         $R_J$&$7\times10^7$&$\unit{m}$\\
         $\Omega$&$1.76\times10^{-4}$&$\unit{rad. s^{-1}}$\\
         $g$&$25$&$\unit{m.s^{-2}}$\\
         $T(0.95R_J)$&$4400$&$\unit{K}$\\
         $T(0.65R_J)$&$11000$&$\unit{K}$
    \end{tabular}
    \tablefoot{Data from \citet{french2012ab,connerney2022new}. Values of spatially varying quantitites are given at $r=0.85R_J$}
    \label{tab:parameters}
\end{table}

\section{Magnetic field strength}\label{app:dimfieldstrength}
To estimate the dimensional magnetic field strength, we assume that the ratio of the mean magnetic to kinetic energies in the simulation is the same as in Jupiter, i.e.
\begin{equation}
    \frac{\bar{E}_b^S}{\bar{E}_u^S}=\frac{e_B}{e_U}.\label{eqn:enratio}
\end{equation}
From Fig.~\ref{fig:energies}, the mean dimensionless kinetic energy in the simulation is $\bar{E}_u^S\approx6\times10^6$, giving an energy density of $\bar{E}_u^S/V=1.64\times10^6$, where $V$ is the fluid volume. We estimate the local kinetic energy density in the semiconvection layer of Jupiter taking density $\rho_0=10^3~\unit{kg m^{-3}}$ and velocity $U=1~\unit{m s^{-1}}$ (at the upper end of the range given by \citet{moore2019time}), to give $e_U=500~\unit{kg m^{-1} s^{-2}}$. The ratio of the planetary energy density to simulation energy is then
\begin{equation}
     Q = \frac{e_U}{\bar{E}_u^S} = 8.33\times10^{-5}~\unit{kg m^{-1} s^{-2}}\label{eqn:Q}
\end{equation}
Using this ratio in equation~\eqref{eqn:enratio}, we obtain an estimate $\tilde{e}_B$ for the local magnetic energy density:
\begin{equation}
    \tilde{e}_B = Q\bar{E}_b^S = 167~\unit{kg m^{-1} s^{-2}}
\end{equation}
For the magnetic field, the simulation field is rescaled by the square root of the ratio $Q$ multiplied by the vacuum permeability (cf. equation~\ref{eqn:energydensities}), giving the dimensional estimate $\tilde{\vec{B}}$ (in Tesla):
\begin{equation}
    \tilde{\vec{B}}=\sqrt{\mu_0Q}\vec{b} = 0.014\vec{b},
\end{equation}
which is plotted in Fig.~\ref{fig:surffield}. 

It should be noted that the value of $Q$ given by equation~\eqref{eqn:Q} depends strongly on the choice of velocity scale $U$, which we have taken to be at the upper end of the range given by \citet{moore2019time}, with $Q\propto U^2$. As such, the redimensionaled field $\tilde{\vec{B}}$ plotted in Fig.~\ref{fig:surffield}(b) is also also on the upper end of the range, so our simulations certainly underestimate $\tilde{\vec{B}}$ compared to the true planetary values in Fig.~\ref{fig:surffield}(c).

\end{appendix}
\end{document}